\documentclass[fleqn,twoside,twocolumn,nofootinbib,showkeys]{revtex4} 
\usepackage[nocpr]{ujp} 

\begin{document}
\title[Graphene Wetting by Methanol or Water]
{GRAPHENE WETTING BY METHANOL OR WATER}
\author{A.G. Barylka}
\affiliation{Kryvyi Rig National University}
\address{54, Gagarin Ave., Kryvyi Rig 50086, Ukraine}
\email{barilkaalena@gmail.com, oks_pol@cabletv.dp.ua}
\author{R.M.~Balabai}%
\affiliation{Kryvyi Rig National University}%
\address{54, Gagarin Ave., Kryvyi Rig 50086, Ukraine}%
\email{barilkaalena@gmail.com, oks_pol@cabletv.dp.ua}
\udk{621.315.592}  \razd{\secvii}

\autorcol{A.G.\hspace*{0.7mm}Barylka, R.M.\hspace*{0.7mm}Balabai}

\setcounter{page}{1049}%

\begin{abstract}
The spatial distributions of the valence-electron density and the
total energy reliefs for water (or methanol) migration on the free
surface of graphene are obtained, by using the electron density
functional and {\it ab initio} pseudopotential methods.\,\,Water and
methanol molecules are found to migrate along the surface of
graphene with an energy relief with barriers and wells.\,\,The
interaction of water molecules located on the opposite sides of the
graphene plane through the regions in the graphene plane with a low
electron density is detected.\,\,The hovering of molecule over the
graphene plane is found to locally change plane's
conductivity.\,\,The estimate of energy costs during the propagation
of adsorbent molecules over the graphene surface testifies to the
graphene hydrophobicity.
\end{abstract}
 \maketitle

\section{Introduction\\ and Statement of the Problem }

Graphene is a two-dimensional allotropic modification of carbon
formed by a monolayer of $sp2$-hyb\-ri\-di\-zed carbon atoms that
are bound by $\sigma$- and $\pi$-bonds into a hexagonal
two-dimensional crystal lattice possessing a unique combination of
properties.\,\,Despite the intense studies of graphene, there are
few reports containing the information on the interaction between
water (methanol) and graphene, which could be important for using
graphene in protective coatings \cite{1,2,3,4,5,6,7}.\,\,The
information concerning the wetting of graphene by water and methanol
can promote the development of various applications.\,\,In
particular, hydrophobic and superhydrophobic surfaces have a low
surface energy and, as a rule, are electric insulators.\,\,A
conformal coating of such surfaces with graphene can result in the
creation of a new class of conducting superhydrophobic surfaces,
which can be applied in various domains \cite{8,9,10,11}.\,\,Bearing
all that in mind, we calculated the energy barriers for water and
methanol migration along the graphene plane in the framework of the
electron density functional and {\it ab initio} pseudopotential
methods.\,\,All calculations were carried out with the help of our
own program \mbox{code \cite{12}.}\looseness=1

\section{Models and Calculation Methods}

All our estimates of static structural properties in terms of the
energy evolution are based on the following
assumptions:\,\,(i)\,\,electrons are in the ground state for every
instant positions of nuclei (the Born--Op\-pen\-hei\-mer adiabatic
app\-ro\-xi\-ma\-tion); (ii)\,\,ma\-ny-par\-tic\-le effects are
estimated in the framework of the local electron density functional
formalism, and (iii)\,\,the so-cal\-led frozen core approximation,
i.e.\,\,pseu\-do\-po\-ten\-ti\-als, is used.\,\,The theory of
pseu\-do\-po\-ten\-ti\-als makes it possible to use the convenient
mathematical apparatus of Fourier transformations, because, owing to
the pseu\-do\-po\-ten\-ti\-al weakness, plane waves can be used as
the basis functions at expanding one-particle electron wave
functions.\,\,Due to the artificial symmetry of examined objects,
the expression for the total energy can be written simpler in the
momentum space.\,\,The total energy per unit cell looks
like\vspace*{-2mm}
\[E_{\rm tot}/\Omega=\sum\limits_{k,G,i}{\left\vert {\Psi_{i}({\bf k}+{\bf G}%
)}\right\vert }^{2}\frac{\hbar^{2}}{2m}({\bf k}+{\bf
G})^{2}\,+\]\vspace*{-7mm}
\[+\,\frac{1}{2}4\pi e^{2}\sum\limits_{G}^{\prime}{\frac{\left\vert
{\rho({\bf G})}\right\vert
^{2}}{{\bf G}^{2}}+\sum\limits_{G}{\varepsilon_{xc}}}({\bf G})\rho^{\ast}%
({\bf G})\,+\]\vspace*{-7mm}
\[+\sum\limits_{G,\tau}^{\prime}{S_{\tau}({\bf G})V_{\tau}^{L}({\bf G})}%
\rho^{\ast}({\bf G})+ \!\!\sum\limits_{k,G,G^{\prime},i,l,\tau}\!\!{S_{\tau}({\bf G}%
-}{\bf G}^{\prime})\,\times\]\vspace*{-7mm}
\[\times\,\Delta V_{l,\tau}^{NL}({\bf k}+{\bf G},{\bf
k}+{\bf G}^{\prime})\Psi_{i}({\bf k}+{\bf G})\Psi_{i}^{\ast}({\bf
k}+{\bf G}^{\prime })\,+\]
\begin{equation}%
\label{eq1}%
 +\bigg\{\!  {\sum\limits_{\tau}{\alpha_{\tau}}}\!\bigg\}\!  \bigg[
{\Omega
^{-1}\sum\limits_{\tau}{Z_{\tau}}}\bigg]  +\Omega^{-1}\gamma_{\mathrm{Ewald}%
},
\end{equation}
where the vector ${\bf k}$ belongs to the first Brillouin zone,
${\bf G}$ is the reciprocal lattice vector, $\Psi_{i}({\bf k}+{\bf
G})$ is the wave function, the subscript $i$ marks occupied states
for a definite ${\bf k}$, $\rho({\bf G})$ is the coefficient in
the expansion series of the valence-electron density, the primed sum
sign means that the term corresponding to $ {\bf G}=0$ is absent,
the subscript $\tau$ enumerates atoms in the unit cell,
$S_{\tau}({\bf G})$ is the structural factor, $V_{\tau}^{L}$ is a
local spherically symmetric pseudo-potential independent of the
quantum orbital number $l$, $\Delta V_{l,\tau}^{NL}$ is the nonlocal
correction to $V_{\tau }^{L}$ dependent on $l$, $Z_{\tau}$ is the
ion charge, and $\gamma _{\mathrm{Ewald}}$ is the Madelung energy of
point-like ions in a uniform negative background.

\begin{figure}%
\vskip1mm
\includegraphics[width=\column]{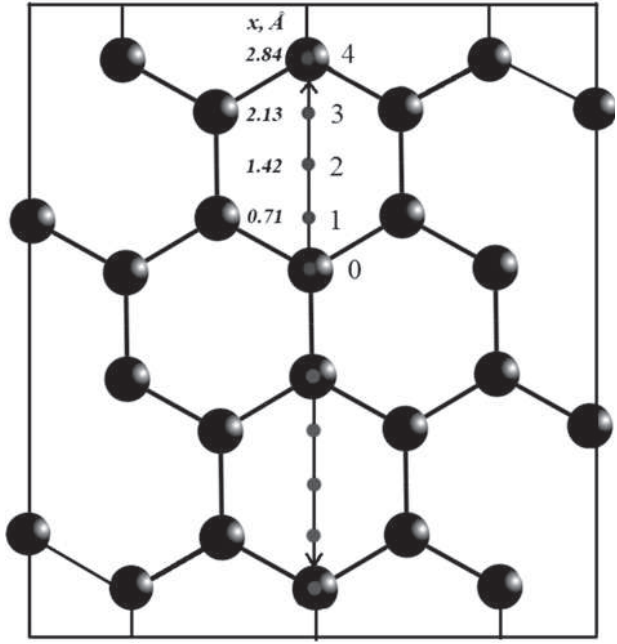}
\vskip-3mm\caption{Schematic diagram of the water (or methanol)
mo\-le\-cu\-le migration on the graphene plane  }
\end{figure}

The coefficients in the Fourier series of the electron charge
density are calculated by the formula
\begin{equation}
\rho({\bf G})=\sum\limits_{i}{\sum\limits_{{\bf
G}^{\prime},\alpha}}\Psi _{i}({\bf k}+{\bf G})\Psi_{i}^{\ast}({\bf
k}+\alpha{\bf G}^{\prime}),
\end{equation}
where $\Psi_{i}({\bf k}+{\bf G})$ are the coefficients of the
expansion in plane waves of the one-particle wave function, which
can be obtained from band-structural calculations, and $\alpha$ is
the operator of symmetric transformations from the point symmetry
group for the unit~cell.

In order to calculate the exchange and correlation energies per
electron, $\varepsilon_{xc}$, we used the Ceperley--Al\-der
approximation parametrized by Perdew and Zun\-ge.\,\,The integration
over ${\bf k}$ was substituted by calculation at $\Gamma$-point.

The capability of graphene to be wetted by water or methanol was
estimated from the energy expenses needed for the adsorbent
molecules to propagate over the surface.\,\,The calculation
procedure was as follows.\,\,First, the adsorption bond length was
determined, i.e.\,\,the distance between water (methanol) molecules
at their migration was chosen.\,\,The choice was associated with the
fact that the minimum distance between a water molecule and the
graphene plane should be equal to the sum of the Bohr radii of the
largest atom in the water molecule and the carbon atom.\,\,The
maximum distance of the molecule from the graphene plane was
determined by the disappearance of the electron density exchange
between the molecule and graphene.\,\,Then, the following scenario
of surface diffusion was used.\,\,Within a diffusion step between
two surface positions, the adsorption bond length was assumed to
remain invariable.\,\,The direction of molecule migration was
selected along the hexagon diagonal (the direction [1100]).\,\,Then
the initial positions of molecules with respect to carbon atoms in
graphene and the displacement step were chosen.\,\,A trajectory of
the described migration is shown in Fig.\,\,1.\,\,The total energy
was calculated for every atomic configuration corresponding to the
elementary step of surface diffusion, thus generating the energy
relief along the
 migration trajectory of a water (or methanol)
 \mbox{molecule.}\looseness=1

Since the calculation algorithm was developed, by assuming the
translational symmetry of the researched atomic system, a supercell
of tetragonal type was created first of all.\,\,Its parameters and
atomic basis depended on the research object.\,\,The atomic basis of
a primitive cell in the artificial lattice reproducing the graphene
plane covered with water molecules from both sides included
30~atoms; and 36 atoms in the case of the plane covered with
methanol molecules.\,\,The account of the translational symmetry was
equivalent to the consideration of an infinite graphene plane
covered with water (or methanol) molecules at a concentration of
8.3\%.

\section{Results of Calculations\\ and Their Discussion}\vspace*{2mm}

With the help of our own program code \cite{12}, we calculated the
total energies of the model atomic system and the spatial
distributions of the valence-electron density.\,\,In addition, the
cross-sections of those spatial distributions were determined for
the $\Gamma$-point in the Brillouin zone of the 3D superlattice.

In Fig.\,\,1, a schematic diagram of water (or me\-tha\-nol)
molecule migration along the graphene plane is shown.\,\,The
direction of water (or methanol) motion is designated by an arrow,
the points mark the positions of molecules, at which the energy of
the studied object was calculated.\,\,Top and bottom points mark the
positions of molecules that move on different sides of the graphene
plane.\,\,Figure~1 also demonstrates the atomic basis of the
primitive cell in the artificial lattice.\,\,The water molecule is
oriented by its oxygen atom toward the graphene plane.

Figures 2 to 4 illustrate variations of the total energy of the
graphene plane during the migration (surface diffusion) of water (or
methanol) molecules along it.\,\,The molecules are separated from
the graphene plane by distances, which are probable for the surface
diffusion processes.\,\,It should be noted that the migration of
molecules occurs simultaneously on the both sides of graphene
plane.\,\,The model trajectory (see Fig.~1) contains a point (this
is position~0, at which the molecule is located above the carbon
atom), when the molecules are the closest to each other.\,\,At the
same time, molecules at position~2 (the atomic hexagon center in the
graphene plane) are the most distant from each other.

Figures 2 and 3, which demonstrate the potential reliefs for water
molecules that migrate in the same direction along the graphene
plane, but at different distances from it, reveal strong
quantitative differences.\,\,In particular, when water molecules
migrate at a short distance from the both sides (lower and upper) of
graphene plane and, so to say, feel each other, the model atomic
system has the highest energy, when the water molecule is located at
the hexagon center (the potential relief has a barrier at this
point).\,\,When water molecules are located in vicinities of carbon
atoms, which screen their interaction, the energy of the atomic
configuration decreases.\,\,%
\begin{figure}%
\vskip1mm
\includegraphics[width=8.1cm]{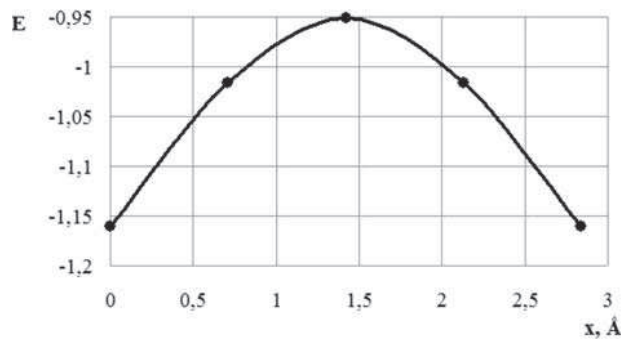}
\vskip-3mm\caption{The potential relief for a water molecule in the
direction [1100] along the graphene surface.\,\,The distance between
the water molecule and graphene equals 1.9~\textrm{\AA }.\,\,The
energy is reckoned in atomic unities per atom, and the distance in
angstr\"{o}ms   }
\end{figure}%
%
\begin{figure}%
\vskip1mm
\includegraphics[width=8.1cm]{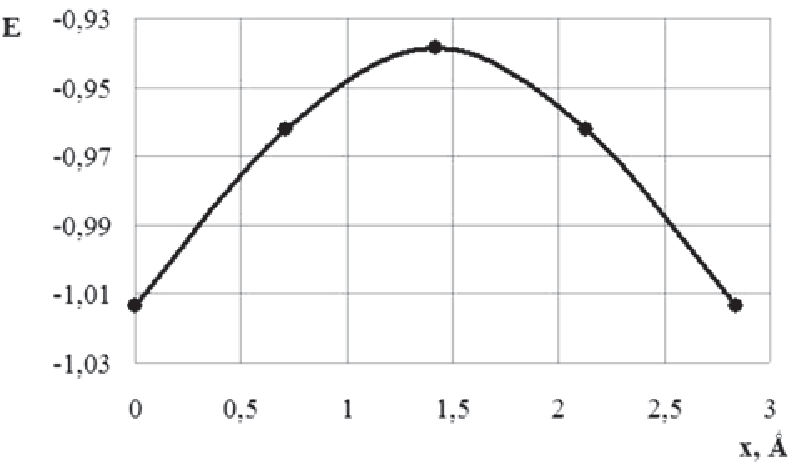}
\vskip-3mm\caption{The same as in Fig.~2, but for the distance
between the water molecule and the graphene plane equal to 2.9~\textrm{\AA }  }
\end{figure}%
%
\begin{figure}[h!]%
\vskip1mm
\includegraphics[width=8.1cm]{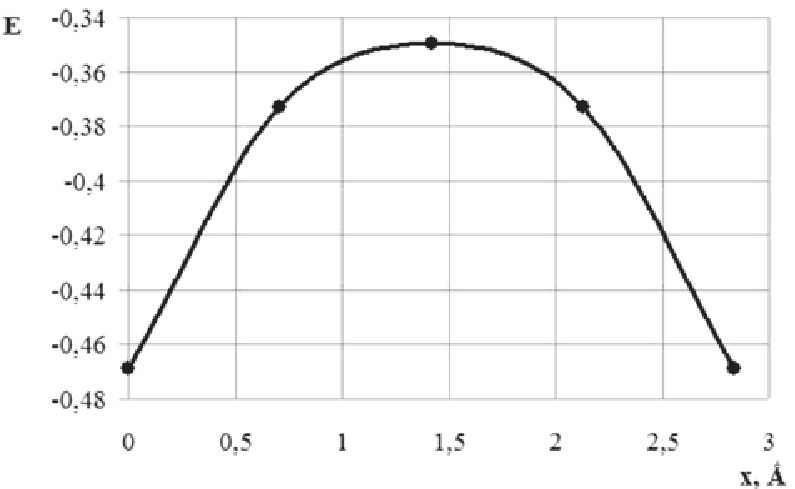}
\vskip-3mm\caption{The same as in Fig.~2, but for the distance
between the water molecule and the graphene plane equal to
3~\textrm{\AA } }\vspace*{-3mm}
\end{figure}%
\begin{figure*}%
\vskip1mm
\includegraphics[width=16.7cm]{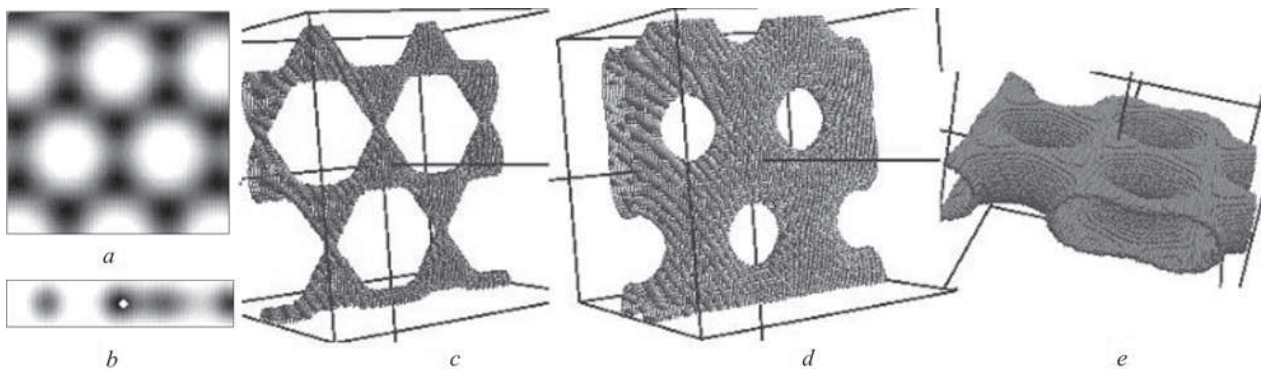}
\vskip-3mm\caption{Distributions of the valence-electron density
near an unconfined graphene plane: longitudinal (\textit{a}) and
transverse (\textit{b}) cross-sections of spatial distribution
through the centers of atoms.\,\,Spatial distributions of the
valence-electron density for isovalues of 0.6--0.7 (\textit{c}) and
0.1--0.2 (\textit{d} and \textit{e}, at different aspect angles)
times the maximum value }
\end{figure*}%
\begin{figure}%
\vskip1mm
\includegraphics[width=\column]{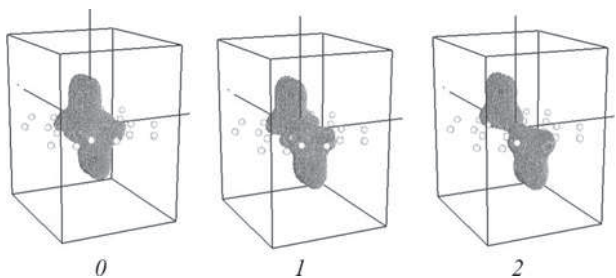}
\vskip-3mm\caption{Spatial distributions of the valence-electron
density near a graphene film covered with migrating water molecules
for isovalues of 0.4--0.5 times the maximum.\,\,Molecules are at a
distance of 1.9~\textrm{\AA } from the graphene plane and move along
the direction [1100].\,\,The numbers of panels correspond to the
numbers of positions in Fig.~1  }\vspace*{-2mm}
\end{figure}%
\begin{figure}%
\vskip1mm
\includegraphics[width=\column]{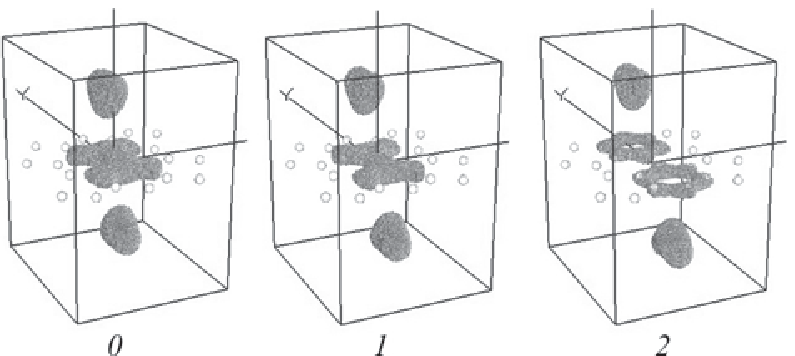}
\vskip-3mm\caption{The same as in Fig.~6, but for the distance
between the water molecule and the graphene plane equal to
2.9~\textrm{\AA }  }\vspace*{-2mm}
\end{figure}%
\begin{figure}%
\vskip1mm
\includegraphics[width=\column]{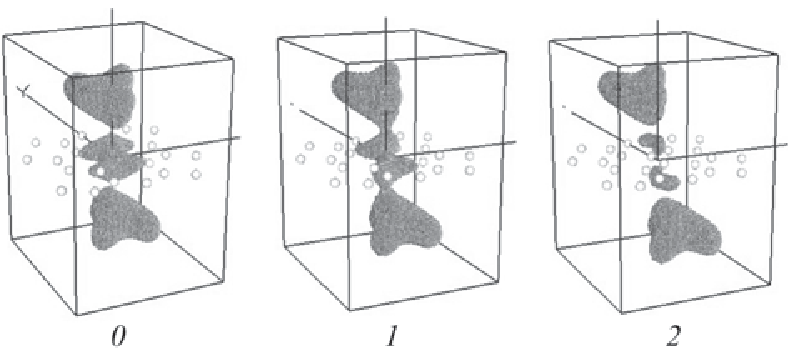}
\vskip-3mm\caption{The same as in Fig.~6, but for the distance
between the water molecule and the graphene plane equal to
3~\textrm{\AA } }\vspace*{-3mm}
\end{figure}%
\begin{figure*}%
\vskip1mm
\includegraphics[width=11cm]{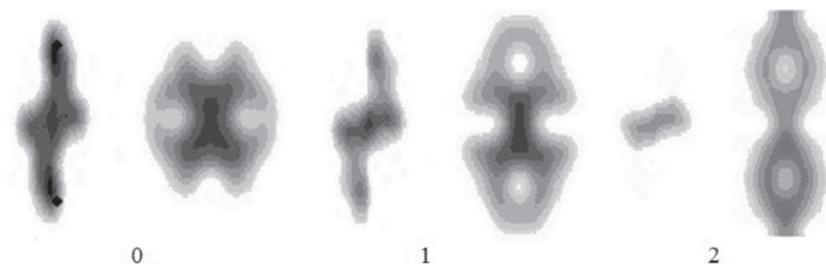}
\vskip-2mm\parbox{11cm}{\caption{Mutually perpendicular
cross-sections of the valence-electron density near the graphene
plane at various positions of a water molecule on the migration
trajectory.\,\,The molecule is at a distance of 1.9~\textrm{\AA }
from the graphene plane.\,\,The numbers of panels correspond to the
numbers of positions in Fig.~1  }}
\end{figure*}%
\begin{figure*}%
\vskip2mm
\includegraphics[width=11cm]{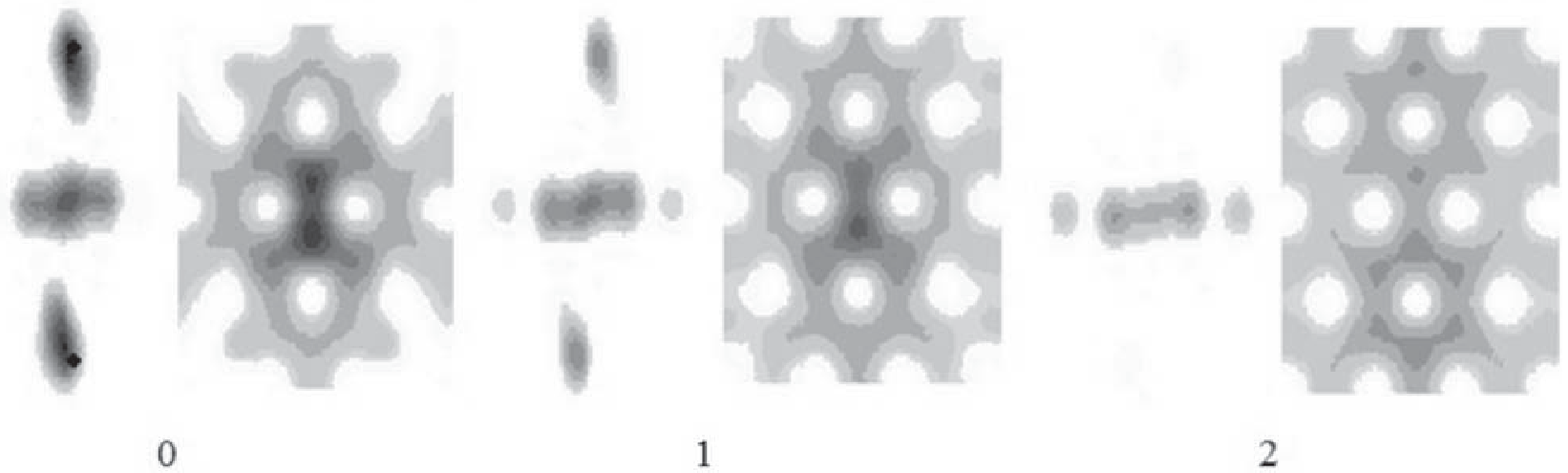}
\vskip-2mm\parbox{11cm}{\caption{The same as in Fig.~9, but for the
distance between the water molecule and the graphene plane equal to
2.9~\textrm{\AA }   }}
\end{figure*}%
\begin{figure*}[!]%
\vskip2mm
\includegraphics[width=11cm]{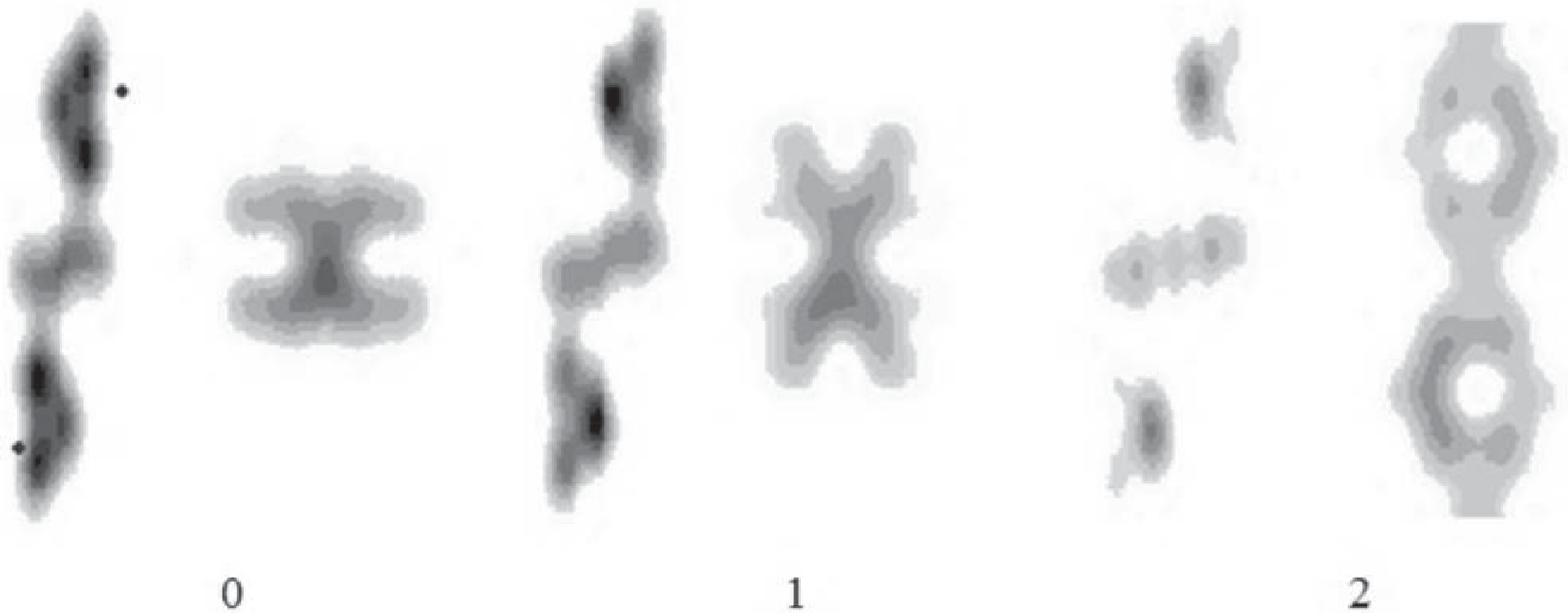}
\vskip-2mm\parbox{11cm}{\caption{The same as in Fig.~9, but for the
distance between the water molecule and the graphene plane equal to
3~\textrm{\AA }  }}\vspace*{-1.0mm}
\end{figure*}%
In other words, water molecules, when approaching the hexagon
vertices, turn out in an energy well.\,\,The energy difference
between the maximum and the minimum of the potential relief amounts
to 0.21~a.u.\,\,per atom.\,\,When water molecules migrate at a long
distance from the both sides (lower and upper) of the graphene plane
and do not feel each other, the energy profile of migration has the
same character.\,\,But the energy difference between the maximum and
the minimum of the potential relief amounts to only 0.07~a.u.\,\,per
atom.\,\,Hence, the height of the energy barrier for the migration
of water molecules on the graphene surface diminished by 66\%.

Attention should be paid to a feature in the electron density
distribution in the case of a free unconfined graphene plane.\,\,At
the centers of atomic hexagons, the own electron density of graphene
equals zero (Fig.\,\,5).\,\,This natural vacuum \textquotedblleft
hole\textquotedblright\ in the graphene plane creates a channel,
through which water molecules located on the different sides of the
graphene plane can interact.\,\,The reality of this interaction is
confirmed by spatial distributions of the electron density and their
cross-sections depicted in Figs.\,\,6--11.\,\,Figures~6--8
illustrate variations in the spatial distribution of the
valence-electron density that accompany the migration of water (or
methanol) molecules along the graphene surface in the direction
[1100], with the molecules being at certain distances from the
graphene plane that are probable for the surface diffusion
processes.\,\,Figures~9--11 demonstrate the cross-sections of those
distributions in mutually perpendicular planes.

Concerning the migration of methanol molecules at a long distance from the
graphene plane (3~\textrm{\AA }), the corresponding energy barrier height
amounts to 0.13~a.u., which is by 50\% larger than in the case of water
molecules that move at the same distance.

Elementary atomic processes running on the gra\-phe\-ne surface
during its wetting by water or alcohol are governed by interactions
between adsorbent molecules and the sur\-fa\-ce.\,\,An exchange of
the electron charge was registered between migrating molecules and
the nearest gra\-phe\-ne atoms, with the exchange intensity
depending on the position of the adsorbed molecule on the surface
(see Fig.~6).\,\,Ano\-ther situation was also observed when the
va\-len\-ce-elect\-ron bridges in the vacuum gap between the
molecules and the gra\-phe\-ne plane were absent, but molecule's
\textquotedblleft sha\-dow\textquotedblright\ with a higher electron
density appeared on the plane under the real molecule (see
Fig.~7).\,\,In other words, the hovering of a molecule above the
gra\-phe\-ne plane locally changed its con\-ductivity.\vspace*{-2mm}

\section{Conclusions}

Using the electron density functional and {\it ab initio}
pseu\-do-potential methods, the distributions of the
va\-len\-ce-electron density and the total energy reliefs for the
water migration on the free graphene surface are
calculated.\,\,Water or methanol molecules migrate over the graphene
surface, which is characterized by an energy relief with barriers
(at the centers of atomic hexagons in the graphene plane) and wells
(in vicinities of carbon atoms).\,\,The barrier height depends on
the distances between the molecules and the graphene
plane.\,\,Interaction through regions with a low electron density in
the graphene plane is registered between water molecules located on
the different sides of the plane.\,\,The hovering of molecules over
the graphene plane is found to locally change plane's conductivity.
The estimation of energy costs during the propagation of adsorbent
molecules on the graphene surface testifies to the phobicity of
graphene with respect to water and alcohol, which is a result of the
non-uniform character of the energy barrier.

\vspace*{-5mm}
\rezume{%
А.Г.\,Барилка, Р.М.\,Балабай}{ЗМОЧУВАННЯ ГРАФЕНУ МЕТАНОЛОМ АБО
ВОДОЮ} {Методами функціонала електронної густини та псевдопотенціалу
із перших принципів отримано розподіли густини валентних електронів
та повні енергії для міграції води (або метанолу) на вільній
поверхні графену. Було встановлено, що міграція молекул води та
метанолу уздовж поверхні графену відбувається з енергетичним
рельєфом, котрому притаманні бар'єри та ями. Зафіксована взаємодія
молекул води, що знаходяться по різні боки від площини графену,
через області малої ймовірності в розподілі електронної густини
графенової площини. Спостерігалося, що зависання молекул над
графеновою площиною локально змінювало її провідність.  Оцінка
енергетичних затрат під час поширення молекул адсорбентів по
поверхні графену продемонструвала його гідрофобність.}

\end{document}